\documentclass[aps,prb,onecolumn,superscriptaddress,showpacs,floatfix, letterpaper]{revtex4-1}
\usepackage[english]{babel}
\usepackage{graphicx}
\usepackage{stmaryrd}
\usepackage{amssymb}
\usepackage{amsfonts}
\usepackage{amsmath}
\usepackage{color}
\usepackage{xspace}

\usepackage{bm}

\begin{document}

\title{Microscopic properties of degradation-free capped GdN thin films studied by Electron Spin Resonance}
\author{Tokuro Shimokawa}
\affiliation{Center for Collaborative Research and Technology Development, Kobe University, 1-1 Rokkodai, Nada, Kobe, Hyogo 657-8501, Japan}
\author{Yohei Fukuoka}
\affiliation{Graduate School of Science, Kobe University, 1-1 Rokkodai, Nada, Kobe, Hyogo 657-8501, Japan}
\author{Masashi Fujisawa}
\affiliation{Research Center for Low Temperature Physics, Tokyo Institute of Technology, 2-12-1 Ohokayama, Meguro-ku, Tokyo 152-8551, Japan}
\author{Weimin Zhang}
\author{Susumu Okubo}
\affiliation{Molecular Photoscience Research Center, Kobe University, 1-1 Rokkodai, Nada, Kobe, Hyogo 657-8501, Japan}
\author{Takahiro Sakurai}
\affiliation{Center for Supports to Research and Education Activities, Kobe University, 1-1 Rokkodai, Nada, Kobe, Hyogo 657-8501, Japan}
\author{Hitoshi Ohta}\thanks{Electronic mail:hohta@kobe-u.ac.jp}
\affiliation{Molecular Photoscience Research Center, Kobe University, 1-1 Rokkodai, Nada, Kobe, Hyogo 657-8501, Japan}
\author{Reddithota Vidyasagar}
\author{Hiroaki Yoshitomi}
\author{Shinya Kitayama} 
\author{Takashi Kita}
\affiliation{Department of Electrical and Electronic Engineering, Graduate School of Engineering, Kobe University, 1-1 Rokkodai, Kobe 657-8501, Japan}

\date{\today}
\begin{abstract}
The microscopic magnetic properties of high-quality GdN thin films have been investigated by electron spin resonance (ESR) and ferromagnetic resonance (FMR) measurements.
Detailed temperature dependence ESR measurements have shown the existence of two ferromagnetic components at lower temperatures which was not clear from the previous magnetization measurements. The temperature, where the resonance shift occurs for the major ferromagnetic component, seems to be consistent with the Curie temperature obtained from the previous magnetization measurement.
On the other hand, the divergence of line width is observed around 57 K for the minor ferromagnetic component.
The magnetic anisotropies of GdN thin films have been obtained by the analysis of FMR angular dependence observed at 4.2 K.
Combining the X-ray diffraction results, the correlation between the magnetic anisotropies and the lattice constants is discussed.
 
\end{abstract}

\maketitle

\section{\label{sec:INTRODUCTION}INTRODUCTION}
Ferromagnetic semiconductors are expected to be a key material for the future spintronics.\cite{Wolf, Senapati}
GdN is one of these ferromagnetic semiconductors and it is particularly interesting due to its partially filled 4f and 5d orbitals with saturation moment of 7 $\mu_{B}$/Gd$^{3+}$.\cite{Dhar}
Therefore, GdN has been the object of a series of theoretical and experimental studies since more than a half century.
\cite{Busch, Sharma, Xiao, Leuenberger, Duan, Gosh, Geshi, Aerts, Lambrecht, Granvile, Li, LiLi, Wachter, Cutler, Gambino}

However, it is well known from a number of studies of bulk GdN in the 60's and 70's \cite{Busch, Wachter, Cutler, Gambino, Duan2, Bush, Schumacher}
that it is very difficult to obtain the high-quality bulk GdN because nitrogen vacancies and oxygen can damage it very easily. 
For example, it was reported that there is a strong decrease in the magnetic moment of bulk GdN even with few percent of oxygen.\cite{Gambino}
The decrease in Curie temperature of bulk GdN was confirmed with a range of oxygen concentration.\cite{Cutler}
Cutler ${\it et \ al}$ also reported that the nitrogen vacancies make the hysteresis effect and the remenance much smaller.\cite{Cutler}
Understanding the properties of "pure" GdN is still challenging because of the difficulty to produce high quality single crystals. 

However, the situation has changed owing to the advanced technology of the thin film synthesis since the 2000s.
\cite{Leuenberger, Granvile, Khazen}
These thin films are capped by AlN on GdN to make surface smooth and to restrict the oxygen contamination.
These early studies of GdN thin film reported that the properties of the GdN thin film, contrary to bulk one, are very sensitive to their epitaxial strain, structural distortion and surface effect for nanocrystalline films.
From the view point of the development in spintronics devices, therefore, it is very important to understand these parameters and how they affect the magnetic and optical properties of the high-quality GdN thin film. 

More recently, H. Yoshitomi ${\it et \ al}$. and R. Vidyasagar ${\it et \ al}$. have studied the optical and magnetic properties in epitaxial AIN/GdN/AIN double heterostructures grown by reactive radio-frequency (rf) sputtering under ultra-pure conditions.\cite{Yoshitomi1, Yoshitomi2, Sagar1, Sagar2, Sagar3}
For example, their high-quality GdN thin film of 95  nm showed the indirect and direct optical transitions, and the considerable size effects of the optical band gap were observed with a decrease in the GdN thickness. They also investigated the saturation magnetization and Curie temperature estimated by Arrott plots as a function of the thickness of GdN.\cite{Granvile, Yoshitomi1}
However, few cases, except for the ferromagnetic resonance (FMR) measurement by K. Khazen ${\it et \ al}$\cite{Khazen}, studied the microscopic magnetic properties of these GdN thin films.

In this study, we investigate the microscopic magnetic properties of high-quality GdN thin films by the detailed temperature dependence of electron spin resonance (ESR), and the angular dependence of FMR at 4.2 K.

\section{\label{sec:EXPERIMENTAL DETAILS}EXPERIMENTAL DETAILS}
We investigate the micro magnetic properties of three GdN samples.
One has the thickness of 95nm whose optical and macro magnetic properties has been investigated by H. Yoshitomi {\it et al}.\cite{Yoshitomi1}
We call this sample ``08GdN" in this paper.  
The other samples has the thicknesses of 29 nm and 97 nm, respectively; we call these two samples ``10GdN" in this paper.
All samples were grown on c-sapphire (0001) substrates at 500 $^\circ$C by reactive radio-frequency(RF) magnetron sputtering\cite{Kishimoto} in an ultrahigh vacuum chamber.
The input RF power was 250 W. 
AlN/GdN/AlN double heterostructures were used to avoid oxidation.\cite{Gambino}
The growth chamber equipped with multitargets for AlN and GdN was separated from the substrate introduction chamber to avoid oxidation of the target when introducing the substrate.
Al(99.99$\%$) and Gd(99.9$\%$) were used as metal targets.
We used an ultrapure (99.9999$\%$) gas mixture of argon and nitrogen for reactive growth.
For the synthesis of 08GdN thin film, the partial pressure ratio of argon and nitrogen was even, and the total sputtering pressure was 5 Pa.
For the syntheses of 10GdN thin films, on the other hand, the partial pressure ration was 9:6, and the total sputtering pressure was 6 Pa for the purpose to decrease the number of nitrogen vacancies.
However, the transmission and absorption spectral measurement showed that the number of free carrier in 10GdN is more than that in 08GdN.
It's not known exactly why the number of nitrogen vacancy in 10GdN is more than that in 08GdN.\cite{doctorYoshi}
The X-ray diffraction measurement showed that the lattice constant along (111) direction for 08GdN sample is $a=0.507$ nm. 
The 29nm and 97nm thin films of 10GdN have $a=0.506$ nm and $a=0.507$ nm, respectively.
These lattice constants are longer than the bulk value $a=0.4998$ nm.
In addition, we also confirmed that the lattice constant along (200) direction is smaller than the bulk value $a=0.250$ nm;
for example, the value for 95 nm thickness of 08GdN is $a=0.249$ nm.
Therefore, our GdN thin films have uniaxial lattice distortion.

Our ESR/FMR measurements were performed by the Bruker X-band ESR spectrometer EXM081 at Center for Supports to Research and Education Activities, Kobe University, with 100kHz field modulation using a TE103 rectangular cavity in the temperature range of 4.2 K to 300 K. 
We show the geometry of the FMR measurements in Fig.~1, which is the same condition as that in Khazen's paper.\cite{Khazen}
The GdN samples lie in the $x$-$z$ plane, and the $y$-axis is parallel to the growth face direction [111] of our thin films.
The out-of plane variation of the external magnetic field is in the $xy$ plane. 
\begin{figure}
 \includegraphics[width=6cm]{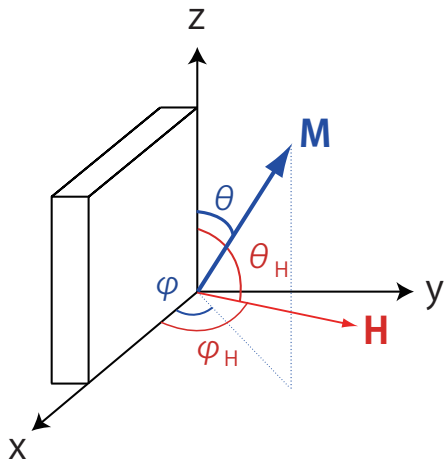}
 \caption{(Color online) The geometry of the FMR measurements. 
${\bf M}$ and ${\bf H}$ are spontaneous magnetization and external magnetic field, respectively.
$\theta$($\theta_{H}$), $\varphi$($\varphi_{H}$) are polar and azimuthal angles for ${\bf M}$(${\bf H}$) vector.   }
 \label{}
\end{figure}

\section{\label{sec:EXPERIMENTAL RESULTS}EXPERIMENTAL RESULTS}
Firstly, we show the temperature dependence of the resonance field.
We successfully obtained the result of Fig.~2 (a) from the spectrum fitting, using the following double lorentzian equation with respect to the ESR spectrum at each temperature.
Here, we would like to point out that the observed ESR signal is the differential curve.
Total intensity $I$ as a function of external magnetic field $H$ is written by 

\begin{eqnarray}
I(H) &=& -16 \frac{I'_{m1}[(H-H_{1})/(\Delta H_{{\rm PP}1}/2)]}{[3+\{ (H-H_{1})/(\Delta H_{{\rm PP}1}/2) \} ^2]^2} \nonumber \\ 
     & & -16 \frac{I'_{m2}[(H-H_{2})/(\Delta H_{{\rm PP}2}/2)]}{[3+\{ (H-H_{2})/(\Delta H_{{\rm PP}2}/2) \} ^2]^2} \nonumber \\ 
     & & +m + m^\prime H,
\end{eqnarray}
where $I'_{mi}$ is the intensity, $H_{i}$ is the resonance field, $\Delta H_{{\rm PP}i}$ is the line width for component $i$ and $m$ and $m^\prime$ are background.
These parameters are determined by the fitting experimental data.
As an example, we show the ESR spectrum and the fitting result at $T=$40 K for 95 nm thickness of 08GdN in the inset of Fig.~2 (a).
Note that the external magnetic field was applied to in the plane ($\varphi_{H}=0$).
Fig.~2 (a) shows the temperature dependence of resonance field for 95 nm thickness of 08GdN and 97nm thickness of 10GdN.
We estimate the $g$-factor $g \sim 1.96$ by using the resonance field value 3524.9 G at the highest temperature 260 K.
This $g$ value is consistent to that of Land$\acute{{\rm e}}$ $g$-factor $g_{L}=2$ of Gd$^{3+}$ whose total orbital angular momentum $L$ is 0. 
At the low temperature region, we can see clearly two kinds of phases not only for 08GdN but also for 10GdN which has the larger number of nitrogen vacancy.
Therefore, the origin of the phase separation is not coming from the nitrogen vacancy.
We have already confirmed the existence of such two kinds of phases in the other GdN samples.\cite{Yoshitomi2}
The Curie temperature ($T_{\rm c}$) for the 95 nm thickness of 08GdN has been reported about 37 K by using Arrott plot analysis\cite{Yoshitomi1}, 
therefore, the shift in the resonance field around 40 K (res. 2 in Fig.~2 (a)) comes from a dominant part of the magnetization of GdN thin film in the ferromagnetic phase. 
In the present study, we also confirmed that the $T_{\rm c}$ value for the 97 nm thicknesses of 10GdN is about 29 K by using Arrott plot analysis. 
The difference of the $T_{\rm c}$ between 08GdN and 10GdN comes from the number of nitrogen vacancy.\cite{Cutler}
Careful observation of Fig.~2 (a) tells us that the resonance shift of 10 GdN (res.~2) begins at lower temperature than that of 08 GdN, and which is consistent that the Curie temperature for 10 GdN is lower than that of 08 GdN.
Therefore, the shift in the resonance field around 30 K (res.~2) for 10GdN also comes from a dominant part of the magnetization in the ferromagnetic phase.
The second shift at higher temperature side ($\sim$70K (res.~1)) is originated from another ferromagnetic phase, which cannot be ascribed to the short-range correlation of spins at $T > T_{\rm c}$
because without the phase separation we cannot observe two ESR's in the intermediate temperature region (25$\sim$57 K).
The high-Tc phase may come from the interface because the contribution of res.~1 to the static magnetization is less than 1\%  for 08GdN at 50 K where the shift of res.~1 is close to the saturation while the shift of res.~2 has just started.\cite{Yoshitomi2} However it requires further investigation.
Here we would like to emphasize that the observation of the two kinds of phases as in the case of Fig.~2 (a) suggests
ESR measurements can easily detect the microscopic properties such as the phase separation which is difficult to be observed by macro measurements.
We also note that it was difficult to measure the resonance field for 10 GdN samples at higher temperature region because the nitrogen vacancy provides carriers at higher temperature which causes the decrease of Q-factor in ESR measurements.

\begin{figure}
 \includegraphics[width=9cm]{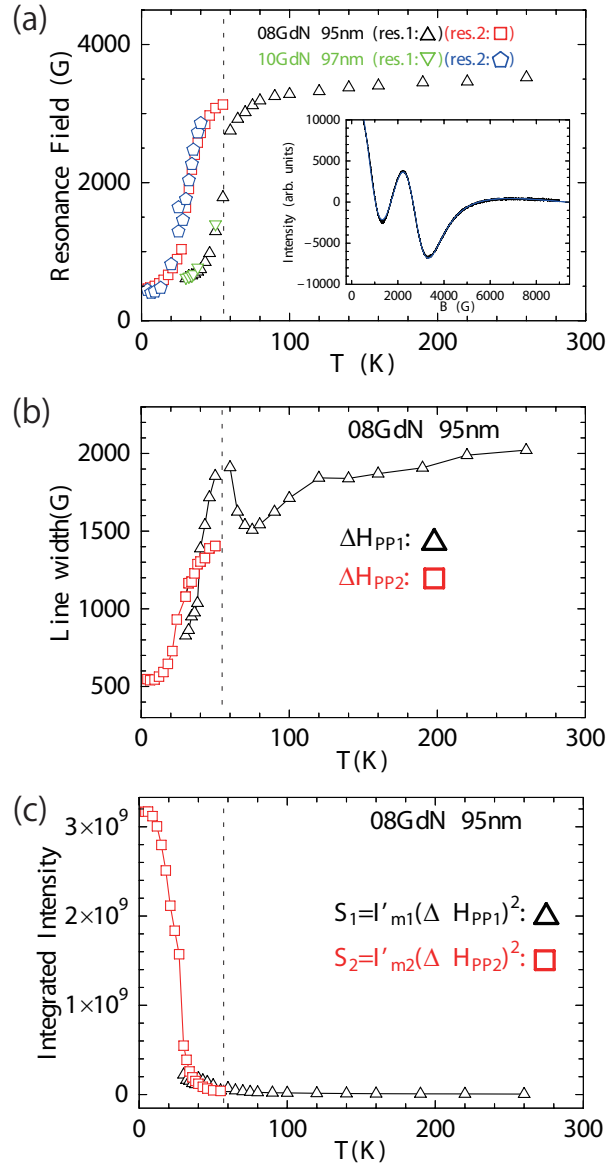}
 \caption{(Color online) (a) Temperature dependences of the resonance field for 95 nm thickness of 08GdN and 97 nm thickness of 10GdN.
 The word of ``res.~1(2)" means the resonance shift at higher (lower) temperature region. 
 The inset is the fitting result at $T=$ 40 K for 95 nm thickness of 08 GdN by using double Lorentzian equation (1).
 (b) and (c) are temperature dependences of the line width and of integrated intensity for 95 nm thickness of 08GdN, respectively. }
 \label{ }
\end{figure}

Next, we investigated the temperature dependence of the line width for 95 nm thickness of 08GdN.
Fig.~2(b) shows the results about the line width.
Owing to ESR measurements, we confirm the decreasing behaviors of the line width $\Delta H_{\rm PP1}$ and  $\Delta H_{\rm PP2}$ 
below 57 K which corresponds to the mid-point of res.~1 resonance shift in Fig.~2(a).
The decreasing behavior with decreasing temperature in the ferromagnetic phase is well known as a typical property in the ferromagnetic region.\cite{Magnetism1}
It is also interesting that the divergence behavior of $\Delta H_{\rm PP1}$ can be observed clearly near at 57 K.
This divergence of the line width indicates the presence of spin fluctuations near $T_{\rm c}$.\cite{Michael} 
The similar increase behavior near above $T_{\rm c}$ has been also confirmed roughly in typical FM thin films.\cite{Michael}
However, no such divergence behavior is observed for $\Delta H_{\rm PP2}$ suggesting the different spin dynamics for res.~1 and res.~2.
We also investigated the temperature dependence of the integrated intensity as shown in Fig.~2(c).
Here, integrated intensity $S_{i}$ for each component $i$ can be calculated by the line width $\Delta H_{{\rm PP}i}$ and intensity $I'_{mi}$.
It is well known in ESR that $S_{i}$ can be estimated by $I'_{mi}(\Delta H_{{\rm PP}i})^2$.
According to this result, we can determine which of two separated phases is smoothly connected to the dominant part in the ferromagnetic ground state.
Therefore, we can say that ``res.~2'' in Fig.~2 (a) and $\Delta H_{\rm PP2}$ in Fig.~2(b) correspond to the dominant part in ferromagnetic state.

Next, we investigated the angular dependence of FMR.
For example, Fig.~3 shows the angular dependence of the FMR for the 08GdN at 4.2K where $\varphi_{H} $ is varied from $-90 ^ \circ $ to $90^ \circ $.
Here, we measured the FMR in the interval of 10 degrees.
We confirm that the resonance field is very sensitive to the angular variation.
Applying the equation (1) to analyze data in Fig.~3 we get the angular dependence of the line width in Fig.~4.
Owing to the closely-spaced angulars we took, we see the peak structures clearly at $\varphi_{H} \sim 90 ^ \circ $ in Fig.~4. 
We also gain the angular dependence of the resonance field.
Fig.~5 presents the angular dependence of the resonance field for our three samples obtained by the Lorentzian fitting.
In order to investigate the film thickness dependence, we add the data for 29 nm thickness of 10GdN sample.
All sample shows that the resonance field was maximized when external magnetic field was applied in the direction of out-of-plane, 
and minimized when in the direction of in-plane.
This result consistent with typical FMR spectra\cite{Ohta}, and with behaviors of Khazen's samples\cite{Khazen}.
A careful observation of Fig.~5 enable us to confirm the resonance field value for 29 nm thickness of 10GdN is lower than those for the other samples. 
This behavior means that the magnetic anisotropy for 29 nm thickness of 10GdN is different from the other samples.

Finally, we analyzed the magnetic anisotropy for our GdN thin film from the angular dependence of the FMR spectra.
We use ``Smit-Beljer formalism"\cite{Smit} for our analysis which is applicable to thin film of cubic symmetry allowing for a possible uniaxial deformation, because our GdN samples also have uniaxial anisotropy in the process of synthesis.
In this case, the energy density $E$ can be written by

\begin{equation}
E=-MH+K_{1}(\alpha_{1}^2 \alpha_{2}^2 + \alpha_{2}^2 \alpha_{3}^2 +\alpha_{3}^2 \alpha_{1}^2) + (2\pi M^2 - K_{\rm u})\alpha_{2}^2,
\end{equation}
which represent the Zeeman interaction, the magnetic anisotropic energy, and the demagnetization energy.
Here, ${\it K}_{1}$ is the fourth order cubic magnetocrystalline anisotropy constant, and ${\it K}_{\rm u}$ is the second order uniaxial anisotropy constant.
$\alpha_{\it i}$ is the direction cosines of the magnetization $M$ relative to the cubic crystal axes, and $H$ is the applied field (See Fig.~1).
In order to analyze the results in Fig.~5, we can generally fix $\theta=\theta_{H}=\pi/2$ in Fig.~1 and use the following equations.
One is the static equilibrium orientation of the magnetization 
\begin{eqnarray}
H {\rm sin}(\varphi_{H}-\varphi) = (4 \pi M - \frac{2 K_{\rm u} }{M}){\rm sin}(\varphi ) {\rm cos}(\varphi ) + \frac{K_{1}}{2M} {\rm sin}(4 \varphi ) , 
\end{eqnarray}
and the other is the resonance field equation

\begin{eqnarray}
(\frac{\omega}{\gamma})^2 &=& [H {\rm cos}(\varphi_{H} - \varphi) + \frac{K_{1}}{M}(2-{\rm sin}^2 2\varphi ) -(4 \pi M - \frac{2 K_{\rm u}}{M}) {\rm sin}^2 \varphi ] \nonumber \\ 
&\times & [H {\rm cos}(\varphi_{H} - \varphi) + \frac{2 K_{1}}{M}{\rm cos} 4\varphi + (4 \pi M - \frac{2 K_{\rm u}}{M}) {\rm cos} 2\varphi ].
\end{eqnarray}
These two fitting equations for FMR measurements are derived from the following three resonance conditions (Smit-Beljers equatons \cite{Smit}):

\begin{eqnarray}
\frac{\partial  E }{\partial  \theta} &=& \frac{\partial  E }{\partial  \varphi }=0, \\
(\frac{\omega}{\gamma})^2 &=& \frac{1}{M^2 {\rm sin}^2\theta} [\frac{\partial ^2 E }{\partial ^2 \theta}\frac{\partial ^2 E }{\partial ^2 \varphi } -(\frac{\partial ^2 E}{\partial \varphi \partial \theta})^2].
\end{eqnarray}
In Fig.~5 the example of the fitting result for 08GdN sample is shown by the black line.
The fitting is rather successfull. Small deviations between the data and the fitting close to $\pm 90^\circ$ may be due to the subtle misalignment ot the 
sample to the applied magnetic field.
We also performed the same fitting to the obtained data for 10GdN 29 nm and 97 nm samples where the fitting lines are not shown in Fig.~5 to avoid the complication in Fig.~5. However, the fittings are also rather successful. 
Table 1 shows our analysis results for magnetic anisotropy constants.
Here the magnetization $M$ for each sample is obtained from the paper.\cite{Sagar3}
These crystal anisotropies $K_{1}$ and $K_{\rm u}$ are much different from the Khazen's results\cite{Khazen}.
Our $K_{1}$ value is almost one third, and $K_{u}$ is two or three times of each value of Khazen's bulk sample, respectively.
The reason of the difference comes from the difference in the crystal growth process.
The Khazen's GdN samples were deposited on (100) oriented Si substrate and these films were polycrystalline.
On the other hand, our samples were grown along to the (111) direction of GdN on c-sapphire (0001) substrates by reactive radio-frequency magnetron sputtering in an ultrahigh vacuum chamber.
More concretely, the lattice constant of the Khazen's extended film was increased 2.4$\%$ uniformly not along to a specific direction, and the $K_{1}$ value of 2.4 $\%$  increased samples are larger than that of Khazen's bulk sample.
The lattice constant along (200) direction of our samples were decreased although the lattice constant along (111) direction were increased.
In fact, the lattice constant along (111) direction for the samples of the 95 nm thickness of 08GdN and 97 nm thickness of 10GdN is is $a=$0.507 nm which is larger than the reported bulk value $a$=0.4998 nm.The lattice constant along (200) direction for 97 nm thickness of 08GdN is a=0.249 nm and this value is smaller than the bulk value of $a=$0.250 nm.
In addition, the coefficient of thermal expansion of our substrate AlN is 4.4 $\times 10^{-6} K^{-1}$ which is larger than that of the Si substrates, 2.4 $\times 10^{-6} K^{-1}$.
Therefore, we can naturally accept the difference between our and Khanzen's samples.
We also note that the $K_{1}$ value for 29 nm thickness of 10GdN is larger than those of our other samples.
The largeness comes from the small lattice constant for 29 nm thickness of 10GdN along (111) direction, $a=$0.506 nm.
In other words, the $K_{1}$ values has the tendency to come close to the bulk value when the lattice constant approaches to the bulk value $a=$0.4998 nm.
We should be careful to the fact that these lattice constant values measured by X-ray diffraction are just average values.
Therefore, we can not discuss about the second order uniaxial anisotropy $K_{\rm u}$ from the view point of the lattice constant because the $K_{\rm u}$ values are mainly affected by the interfacial surface of crystal.
It is naturally expected that our $K_{\rm u}$ values are more sensitive than Khazen's sample because Khazen's films are polycrystalline.
This is the origin that our $K_{\rm u}$ values are larger than those of Khazen's.
The $K_{\rm u}$ value of 95 nm thickness for 08 GdN is slightly larger than that of 97 nm thickness for 10GdN. 
We speculate that it may come from the nitrogen vacancy, that means, the strain at the interfacial surface for 10GdN samples was relaxed by the large number of nitrogen vacancy.
We also comment about the characteristic which our thinner sample of 29 nm thickness for 10 GdN has the largest $K_{\rm u}$ value.
It is characteristic of ferromagnetic thin films that the thinner thickness sample has the larger value of $K_{\rm u}$.
This behavior is well known theoretically and experimentally, for example, Fe/MgO multilayered films.\cite{Ohta}
According to these obtained results, the FMR analysis is very useful to obtain the microscopic properties.

\begin{figure}
 \includegraphics[width=6cm]{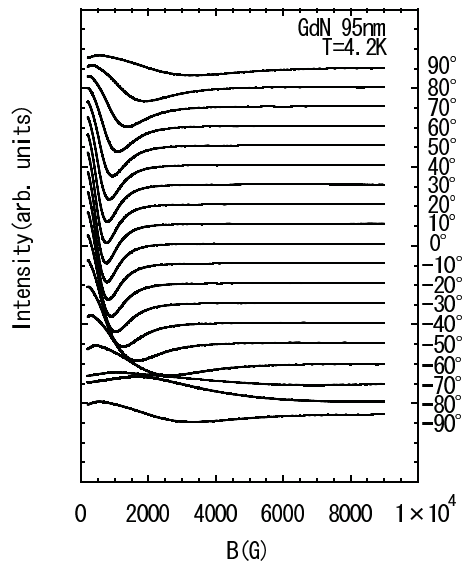}
 \caption{Angular dependence of FMR measurement at $T=$4.2 K for 95 nm thickness of 08GdN.}
 \label{}
\end{figure}

\begin{figure}[h]
 \includegraphics[width=7.5cm]{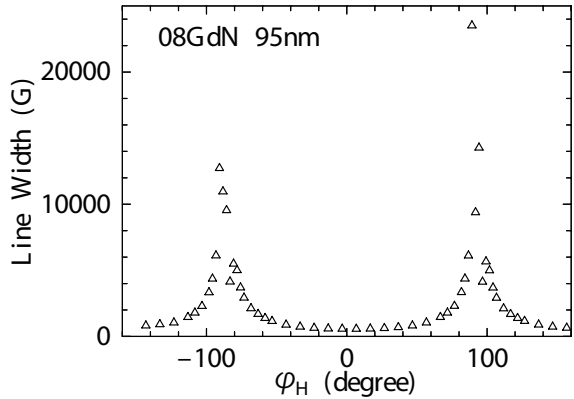}
 \caption{Angular dependence of line width at $T=$4.2 K for 95 nm thickness of 08GdN.   }
 \label{}
\end{figure}

\begin{figure}[h]
 \includegraphics[width=7.5cm]{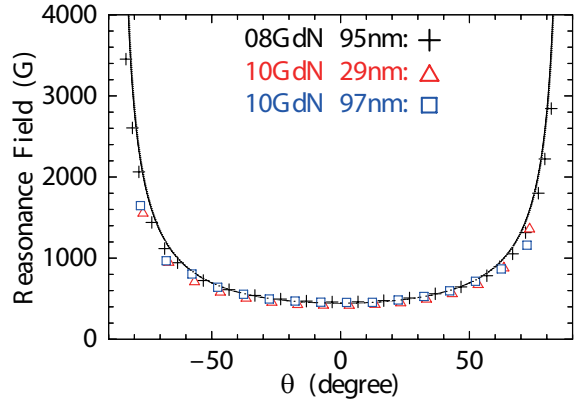}
 \caption{(Color online) Angular dependences of resonance field of FMR for our all samples at $T=4.2$ K.
The external magnetic field was applied out-of-plane at $-90^\circ$ and $90^\circ$. The black line is the fitting result for 08GdN sample.  }
 \label{}
\end{figure}

\newpage
\begin{table}[h]
  \begin{tabular}{l|c|c|c|c|c|c|c} \hline
                      & 4$\pi M$ (Oe) & 2$K_{u}/M$ (Oe) & 2$K_{1}/M$ (Oe) & $K_{\rm u}$ (erg/cm$^3$) & $K_{1}$ (erg/cm$^3$) & T(K) \\ \hline 
    08GdN 95nm & 24167 & 11167 & 404 & 1.07$\times 10^7$ & 3.88$\times 10^5$ & 4.2 &in this paper \\ \hline
    10GdN 29nm & 28660 & 14760 & 394 & 1.68$\times 10^7$ & 4.49$\times 10^5$ & 4.2 &in this paper\\
    10GdN 97nm & 24027 & 10527 & 380 & 1.01$\times 10^7$ & 3.63$\times 10^5$ & 4.2 &in this paper\\ \hline
    Bulk film& 22220  &   5759 & 1292&5.09$\times 10^6$ & 1.14$\times 10^6$ & 4.0  & ref.~ 23\\
    2.4 $\%$ extended film& 15620& 2897 & 2252 & 1.8 $\times 10^6$ & 1.4 $\times 10^6$ & 4.0 & ref.~ 23 \\ \hline  
  \end{tabular}
  \caption{Spontaneous magnetization $M$, the fourth order cubic magnetocrystalline anisotropy constant $K_{1}$ and the second order uniaxial anisotropy constant $K_{\rm u}$ for our samples and Khazen's.}
\end{table}
%\begin{table}[h]
%  \begin{tabular}{l c c l|} \hline
%    AlN          & 4.4 $\times 10^{-6}$ & this work                     \\ \hline
%    Si           & 2.4 $\times 10^{-6}$ & ref.~[23]  \\ \hline
%    Al$_{2}$O$_{3}$  & 7.0 $\times 10^{-6}$ &   this work             \\ \hline
%  \end{tabular}
%  \caption{}
%\end{table}

\section{\label{sec:CONCLUSION}CONCLUSION}
We have investigated microscopic properties of high-quality GdN thin films.
Detailed temperature dependence ESR measurements have been performed for the first time and they showed the existence of two ferromagnetic components at lower temperatures. It also showed that the temperature, where the resonance shift occurs for the major ferromagnetic component, seems to be consistent with the Curie temperature obtained from the previous magnetization measurement.
On the other hand, the divergence of line width is observed around 57 K for the minor ferromagnetic component.
We have also determined  the fourth order cubic magnetocrystalline and second order uniaxial anisotropies of our GdN samples from the angular dependence of FMR measurements observed at 4.2 K.
Our analysis by Smit-Beljer formalism have clarified that the cubic anisotropy is very sensitive to the lattice constant of thin film and the uniaxial anisotropy values 
depend on the thickness of thin films strongly.

%\section{\label{sec:ACKNOWLEDGMENTS}ACKNOWLEDGMENTS}
%We thank Dr. R.~Vidyasagar for fruitful discussions.

%

\end{document}